\documentclass[epj]{webofc}
\usepackage[varg]{txfonts}   % Web of Conferences font

\def\Pom{{\bf I\!P}}
\woctitle{MESON2016 - the 14$^\textrm{th}$ International Workshop on Meson Production, Properties and Interaction}
\begin{document}
\selectlanguage{english}
\title{Photoproduction of vector mesons: \\from ultraperipheral to semi-central heavy ion collisions}

% insert email only for speaker/presenter
\author{Mariola K{\l}usek-Gawenda\inst{1}\fnsep\thanks{\email{mariola.klusek@ifj.edu.pl},
} \and
        Antoni Szczurek\inst{1,2} 
% comment out the next line if not needed
%       \\for the XXXXX Collaboration
}

\institute{Institute of Nuclear Physics, Polish Academy of Sciences, Krak\'ow, Poland \and
           University of Rzesz\'ow, Rzesz\'ow, Poland
          }

\abstract{%Do not break line here!
We discuss nuclear cross sections for $AA \to AAV$ and $AA \to AAVV$
reactions with one or two vector mesons in the final state. 
Our analysis is done in the impact parameter space equivalent photon 
approximation. 
This approach allows to consider the above processes taking into
account distance between colliding nuclei. We consider both
ultraperipheral and semi-central collisions.
We are a first group which undertook a study of single $J/\psi$ photoproduction 
for different centrality bins. We show that one can describe
new ALICE experimental data by including geometrical effects of collisions in the flux factor.

Next, total and differential cross section for double-scattering
mechanism in the exclusive $AA \to AAVV$ reaction in ultrarelativistic
ultraperipheral heavy ion collisions is presented. In this context we
consider double photoproduction and photon-photon processes. 
Simultaneously, we get very good agreement of our results with STAR (RHIC), 
CMS and ALICE (LHC) experimental data for single $\rho^0$ and $J/\psi$
vector meson production. A comparison of our predictions for exclusive 
four charged pions production is also presented.

}
\maketitle
%

%----------------------------
\section{Introduction}
%----------------------------

In last years there was an interest in calculating cross sections 
for exclusive production of vector mesons in ultrarelativistic heavy ion
collisions \cite{FSZ2002,BKN2005,GM2006,HBT2005,Baltz2007,KSS2009,BCKGSS2013,KGS2014,FGSZ2016,KGS2016}.
Purely ultraperipheral collisions (UPC) of heavy ions mean
the situation when the nuclei do not collide, however, an emission of extra neutrons 
caused by additional purely electromagnetic interactions could be included.
In theoretical calculations UPC means in practice $b > R_A + R_B$ 
(sum of nuclear radii). All the calculations must be therefore performed
in the impact parameter space to include this condition.

We discuss an importance of photoproduction process in the context 
of $J/\psi$ production in less peripheral collisions. Then, at high energies,
nuclei can collide and break apart producing quark-gluon plasma.
At low transverse momenta and small multiplicities the ALICE Collaboration 
\cite{ALICE2012} observed an enhancement of nuclear modification factor
at small transverse momenta $R_{AA} > 1$.

Nuclear cross sections are calculated with the help of 
equivalent photon approximation (EPA) in the impact parameter space. 
We include realistic form factor of nuclei which is Fourier transform
of the charge distribution in nucleus. The differences of the cross
sections with realistic and monopole form factor can be found 
e.g. in Ref. \cite{KGS2010}.

%--------------------------------------------
\section{Some theoretical aspect}
%--------------------------------------------

\begin{figure}[ht]
\centering
\includegraphics[scale=0.4]{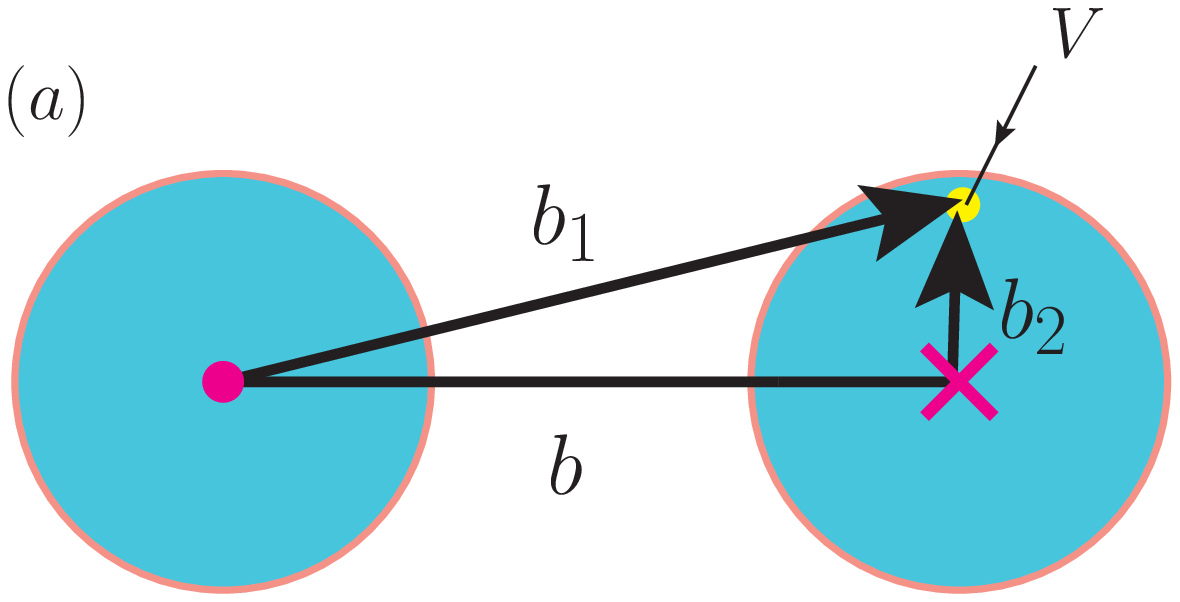}
\hspace{2.cm}
\includegraphics[scale=0.4]{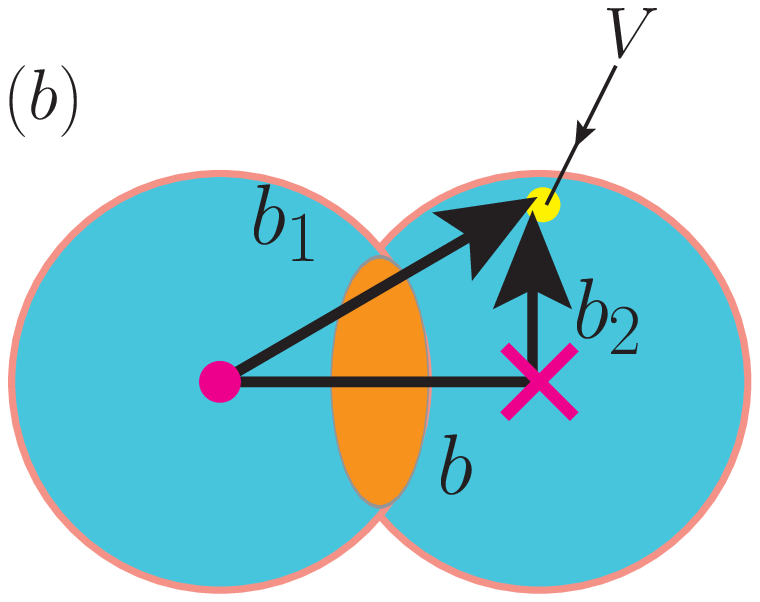}
\caption{Impact parameter picture of the collision and 
the production of the vector meson for ultraperipheral
(left panel) and for semi-central (right panel) collisions. 
It is assumed here that the first nucleus is the emitter of the photon 
which rescatters then in the second nucleus being a rescattering medium.}
\label{fig_per_cent}       % Give a unique label
\end{figure}

Fig. \ref{fig_per_cent} depicts the difference between ultraperipheral
and semi-central collisions of heavy ions. The left panel shows UPC for
production of vector mesons.
The two-dimensional vector $\vec{b}$ is a distance between colliding nuclei, 
$\vec{b_1}$ is a distance between the photon position and the middle of 
the first (emitter) nucleus  and $\vec{b_2}$ is a distance between 
the photon position and the middle of the second (medium) nucleus.
The production of the vector meson may occur provided the photon
(hadronic fluctuation) hits the second nucleus. The hatched area of overlapping nuclei
in the right panel (semi-central collision) of the figure represents the area in the impact
parameter space for which some absorption may be expected.
We include the effect of the ”absorption” by modifying effective photon fluxes in the
impact parameter space by imposing several additional geometrical conditions on impact
parameters (between photon and nuclei and/or between colliding nuclei).

\begin{figure}[!h]
\centering
\includegraphics[scale=0.25]{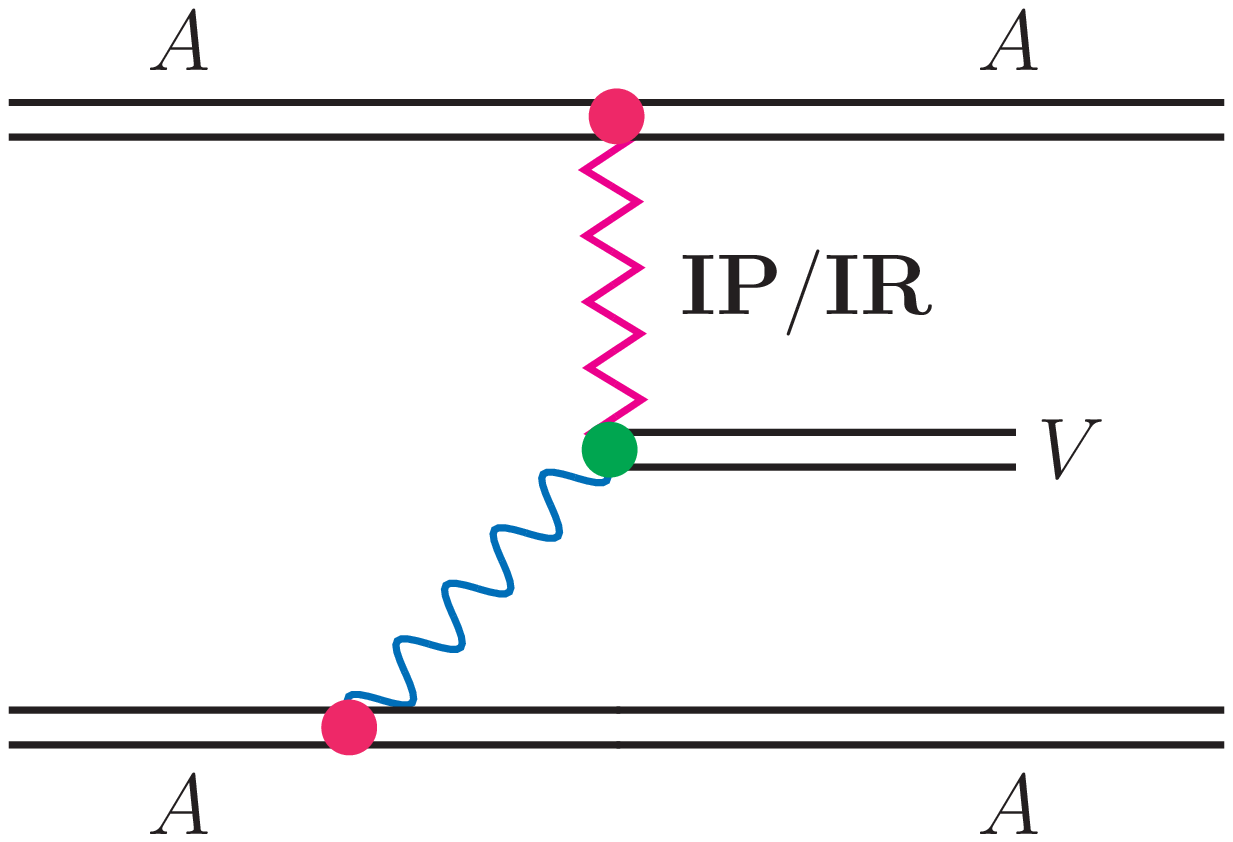}
\includegraphics[scale=0.25]{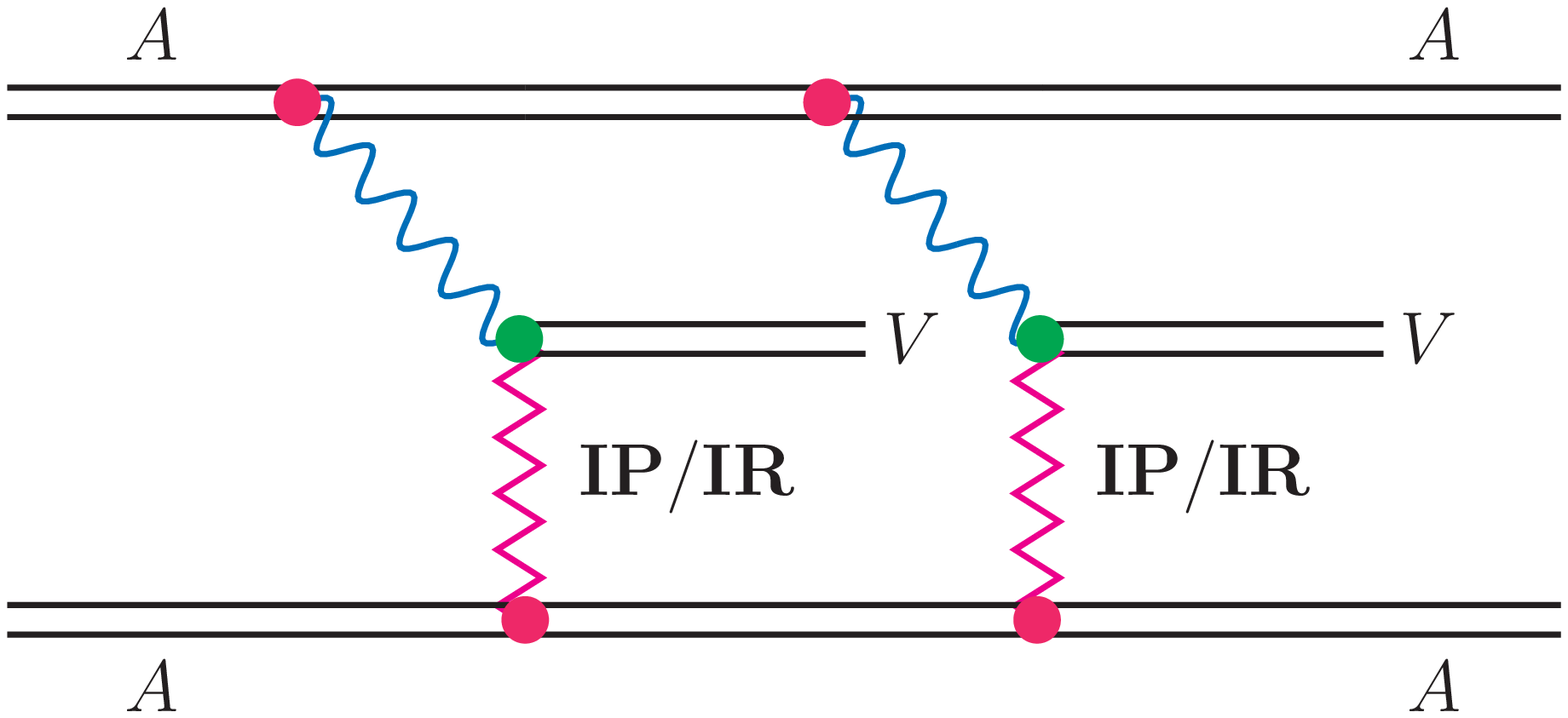}
\includegraphics[scale=0.28]{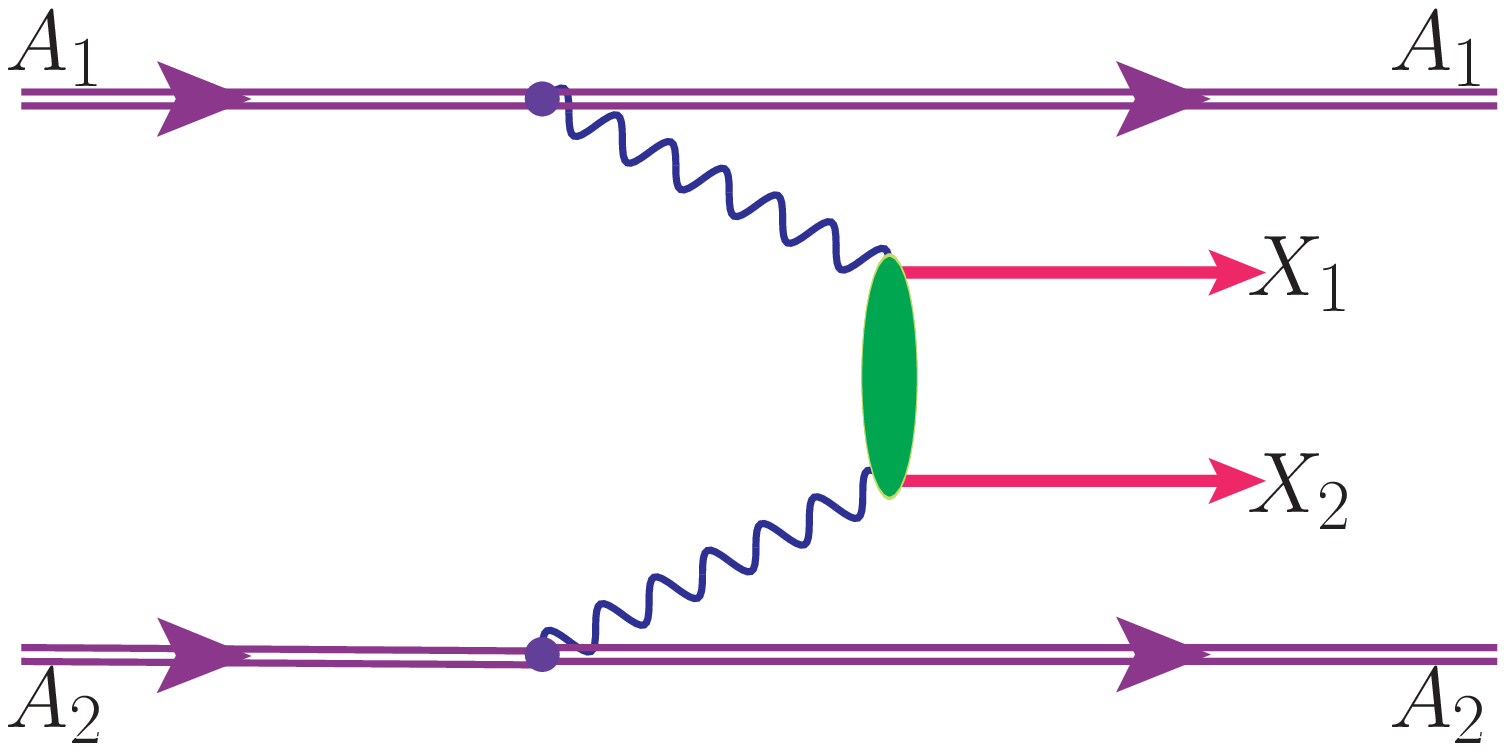}
\caption{Single vector meson production by photon-Pomeron (or Pomeron-photon)
fusion (left panel) and double vector meson production 
for double-scattering mechanism (middle panel) and for $\gamma\gamma$ fusion 
(right panel).}
\label{fig_photoproduction}       % Give a unique label
\end{figure}

The left panel of Fig. \ref{fig_photoproduction} shows a single vector meson production mechanism.
Photon emitted from a nucleus fluctuates into hadronic or
quark-antiquark components and converts into an on-shell meson. 
The differential cross section for this mechanism can be written as 
\begin{equation}
\frac{\mathrm{d} \sigma_{A_1A_2 \to A_1A_2V}}{\mathrm{d}^2 b \mathrm{d}y} = 
\frac{\mathrm{d}P_{\gamma \Pom}\left(b,y\right)}{\mathrm{d}y} + 
\frac{\mathrm{d}P_{\Pom \gamma}\left(b,y\right)}{\mathrm{d}y} \;.
\end{equation}
$P_{\gamma \Pom}(y,b)$/$P_{\Pom \gamma}(y,b)$ are the probability 
densities for producing a vector meson at rapidity $y$ for fixed impact parameter $b$ 
(distance between two colliding nuclei) of the heavy ion collision. 
Each probability is the convolution of a flux of
equivalent photon and the $\gamma A \to V A$ cross section.

The middle panel of Fig. \ref{fig_photoproduction} illustrates one of four 
mechanisms of the double-scattering mechanism (in fact, we take into account 
all four combinations of $\gamma \Pom$ exchanges: 
$\gamma \Pom - \gamma \Pom$, $\gamma \Pom - \Pom \gamma$,
$\Pom \gamma - \gamma \Pom$, $\Pom \gamma - \Pom \gamma$). 
Having a formalism for the calculation of single-vector-meson production, 
one can use it to calculate cross section for double-scattering mechanisms 
of two-vector-meson production. The cross section for the double vector meson
photoproduction is expressed with the help of probability density of single meson 
production
\begin{equation}
\frac{\mathrm{d} \sigma_{A_1A_2 \to A_1A_2VV}}{\mathrm{d}y_1 \mathrm{d}y_2} = 
\frac{1}{2} \int \mathrm{d}^2b \left[
\left(\frac{\mathrm{d}P_{\gamma \Pom}\left(b,y_1\right)}{\mathrm{d}y_1} + 
\frac{\mathrm{d}P_{\Pom \gamma}\left(b,y_1\right)}{\mathrm{d}y_1}\right) \times
\left(\frac{\mathrm{d}P_{\gamma \Pom}\left(b,y_2\right)}{\mathrm{d}y_2} + 
\frac{\mathrm{d}P_{\Pom \gamma}\left(b,y_2\right)}{\mathrm{d}y_2}\right) \right] \;.
\end{equation}
For completeness, in the right panel of Fig.\ref{fig_photoproduction}, 
we show Feynman diagram for pair of vector meson production in the $\gamma\gamma$ fusion.
A formula for the total cross section for this mechanism can be written
in a somewhat simplified way as
\begin{equation}
\sigma_{A_1A_2 \to A_1A_2VV} = \int \mathrm{d}\omega_1 \mathrm{d}\omega_2 
n(\omega_1) n(\omega_2) \sigma_{\gamma\gamma \to VV} \; ,
\label{eq_ggfusion}
\end{equation} 
where $\omega_{1/2}$ is the photon energy from first/second nucleus.
More details referring to Eq. \ref{eq_ggfusion} can be found in Ref. \cite{KGS2014}.

%----------------------------------------------------------------------------
\section{Theoretical results versus experimental data}
%----------------------------------------------------------------------------

\begin{figure}[h!]
\centering
\includegraphics[scale=0.3]{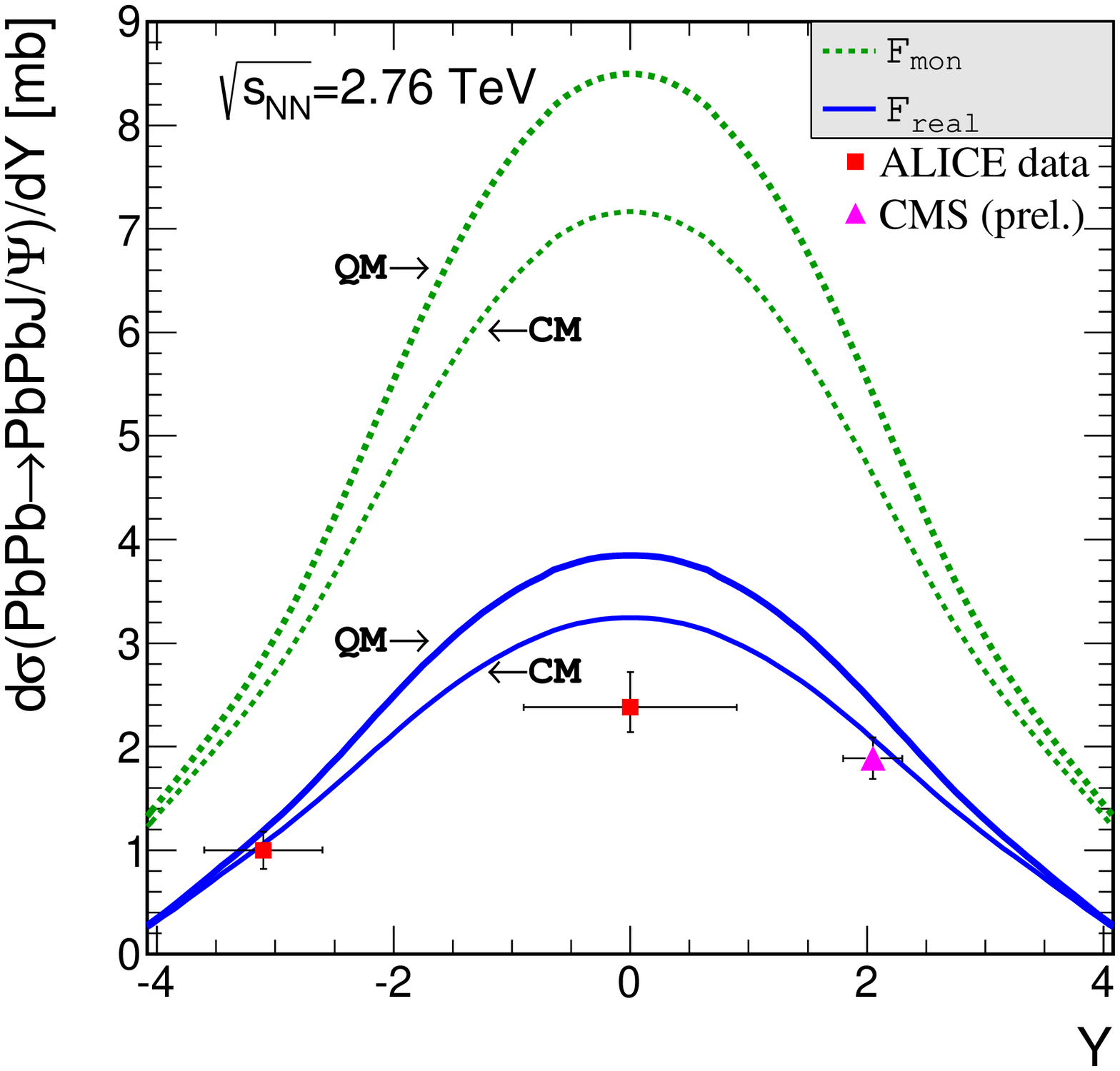}
\includegraphics[scale=0.3]{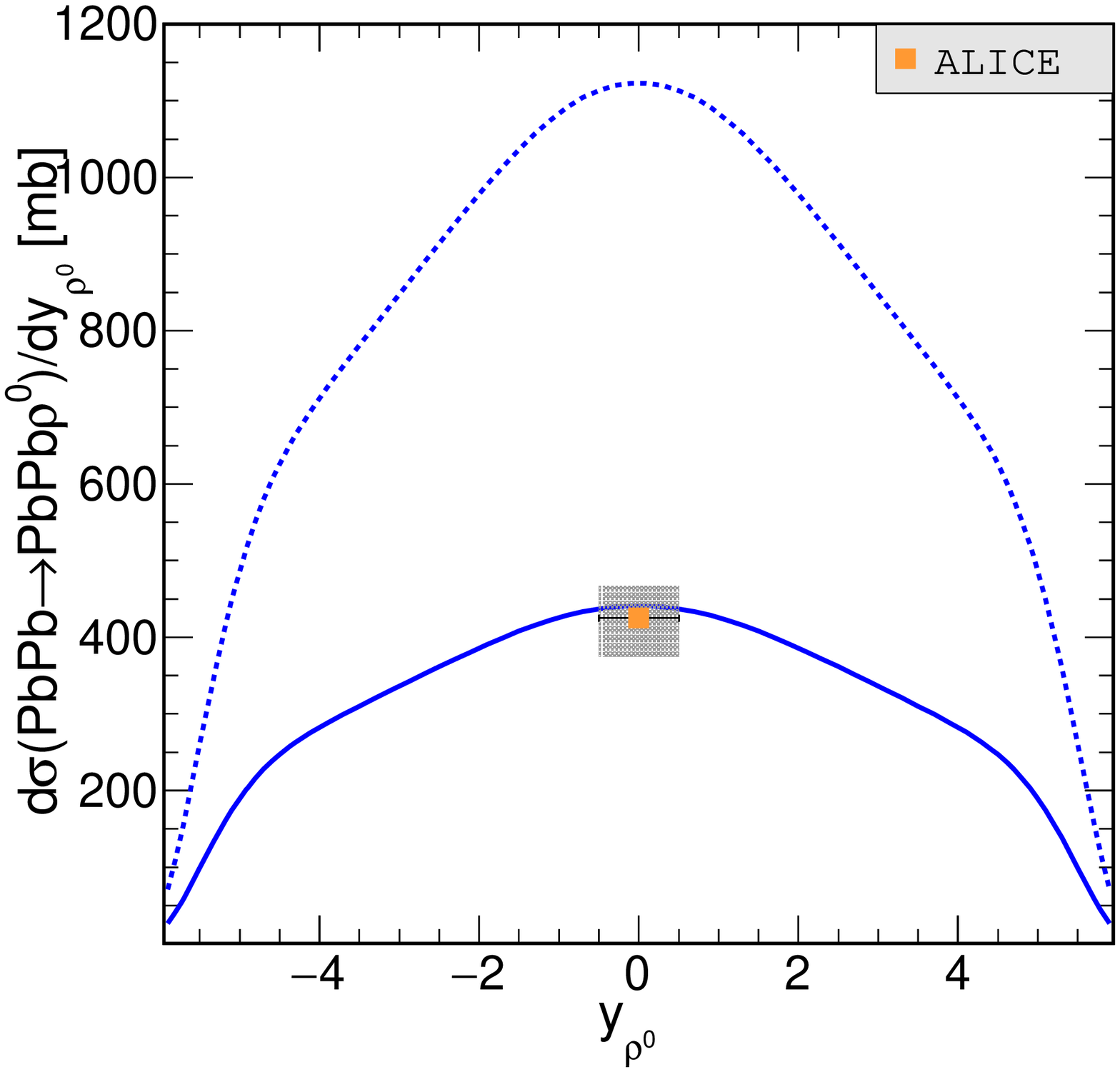}
\caption{Left panel - differential cross section for coherent production of $J/\psi$ meson 
in UPC as a function of rapidity of the $J/\psi$ meson compared with 
the ALICE and CMS data points. Results are shown for
both realistic and monopole form factor.
Right panel - differential cross section for single $\rho^0$ photoproduction
in UPC as a function of meson rapidity together with the ALICE experimental
data point.}
\label{fig_dsig_dy_jpsi_rho}       % Give a unique label
\end{figure}

We start from purely ultraperipheral collisions. 
Fig. \ref{fig_dsig_dy_jpsi_rho} shows results for coherent 
$J/\psi$ (left panel) and $\rho^0$ (right panel) photoproduction 
in the lead-lead UPC at the LHC energy. 
We show our results for classical and quantal rescattering 
(more details in Ref.~\cite{KGS2016}) in the nucleus-medium and for
realistic and monopole nuclear form factors. 
The result with the monopole form factor overestimates
the ALICE \cite{ALICE_jpsi_1,ALICE_jpsi_2} and CMS \cite{CMS_jpsi} data. 
The result with the quantal rescattering is a little larger
than that for the classical rescattering. 
However, this difference is much smaller for $J/\psi$ production than for the
photoproduction of $\rho^0$ meson. the ALICE data point (Ref.~\cite{ALICE_rho0})
for $\rho^0$ meson production (right panel)
is much better described by the classical rescattering calculations (lower line).

\begin{figure}[ht]
\centering
\includegraphics[scale=0.3]{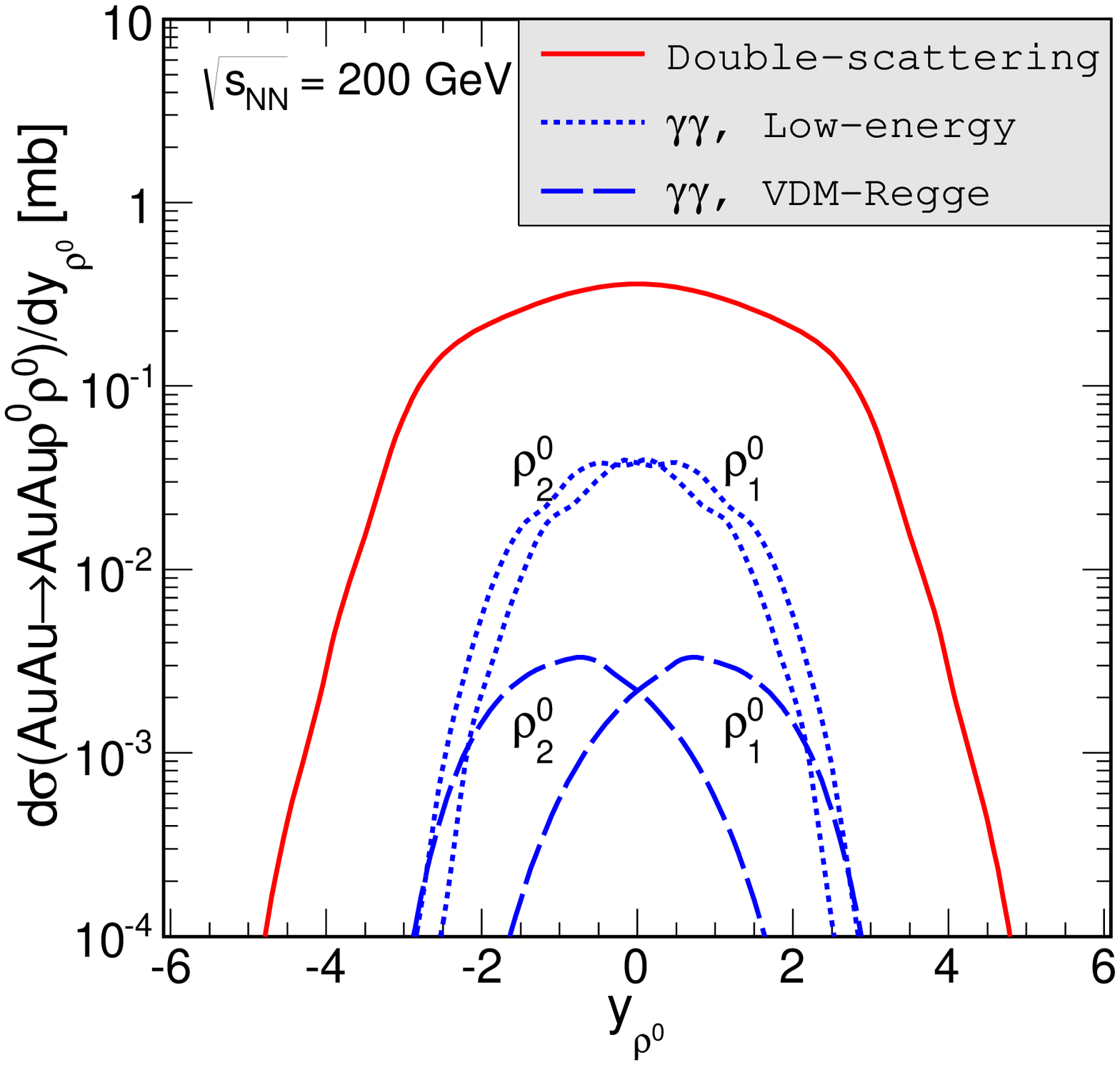}
\includegraphics[scale=0.3]{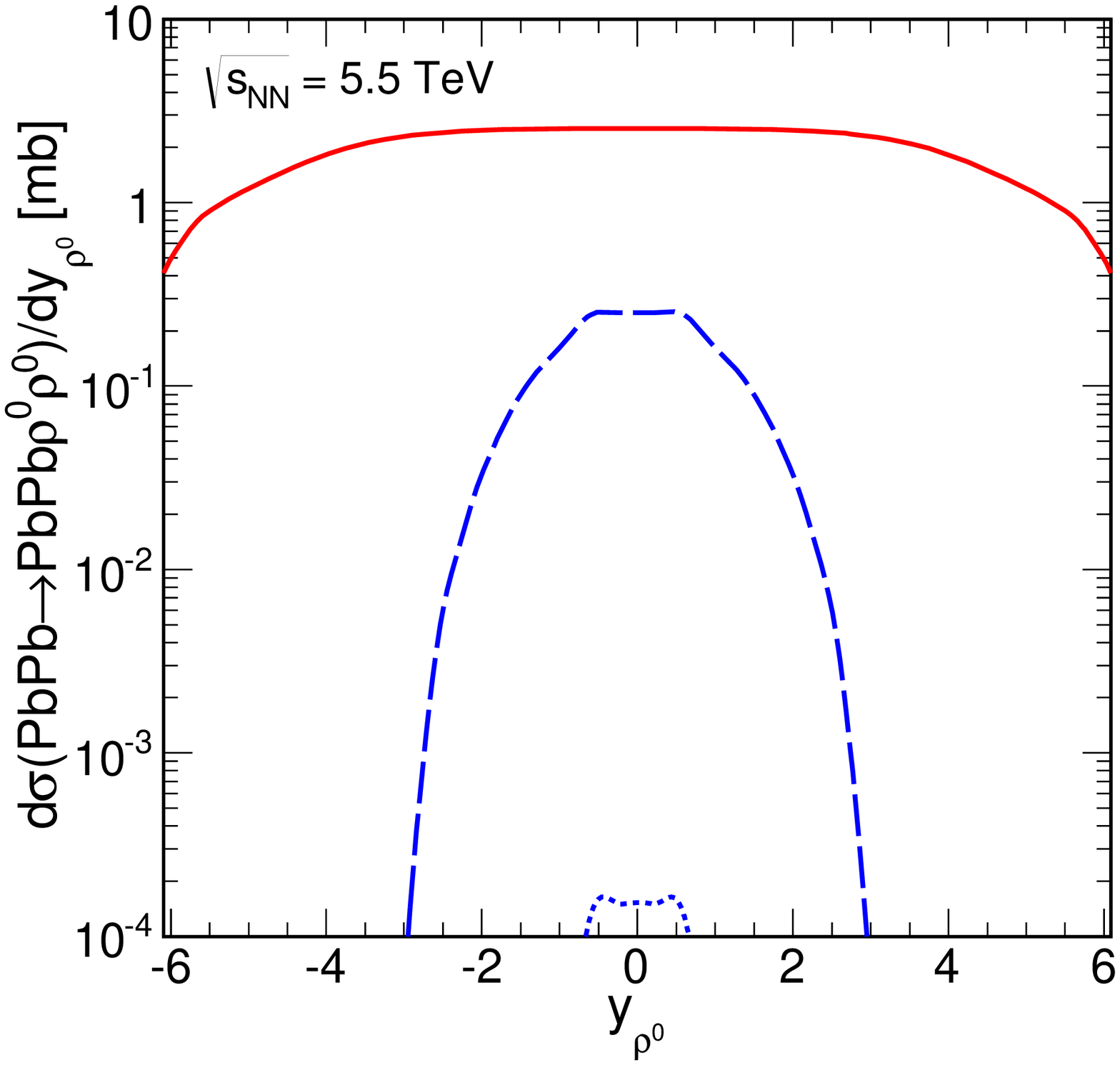}
\caption{Rapidity distribution of one of the $\rho^0$ meson produced 
in the double-scattering mechanism and in the $\gamma\gamma$ fusion 
at the RHIC (left panel) and at the LHC (right panel) energy.}
\label{fig_dsig_dy_rho}       % Give a unique label
\end{figure}

Fig. \ref{fig_dsig_dy_rho} shows differential cross section for 
the $AA \to AA \rho^0 \rho^0$ reaction
as a function of one $\rho^0$ meson rapidity
and a comparison of the double-scattering and $\gamma\gamma$ fusion contributions
at RHIC (left panel) and at LHC (right panel) energies. 
The double-scattering component dominates 
over the $\gamma\gamma$ one. In addition, the distribution
in rapidity for the RHIC energy is much narrower than that for LHC energy.
The high-energy (VDM-Regge) component of the
elementary cross section dominates over the low-energy bump contribution
at the LHC energy.
This is caused by the fact that at the higher center-of-mass energy 
the higher values of two-meson invariant mass becomes more important 
which corresponds to larger values of particle rapidity. 
Both at the RHIC and LHC energy, the contribution
of the $\gamma\gamma$ fusion is one order of magnitude smaller 
than that of the double-scattering mechanism.

\begin{figure}[ht]
\centering
\includegraphics[scale=0.3]{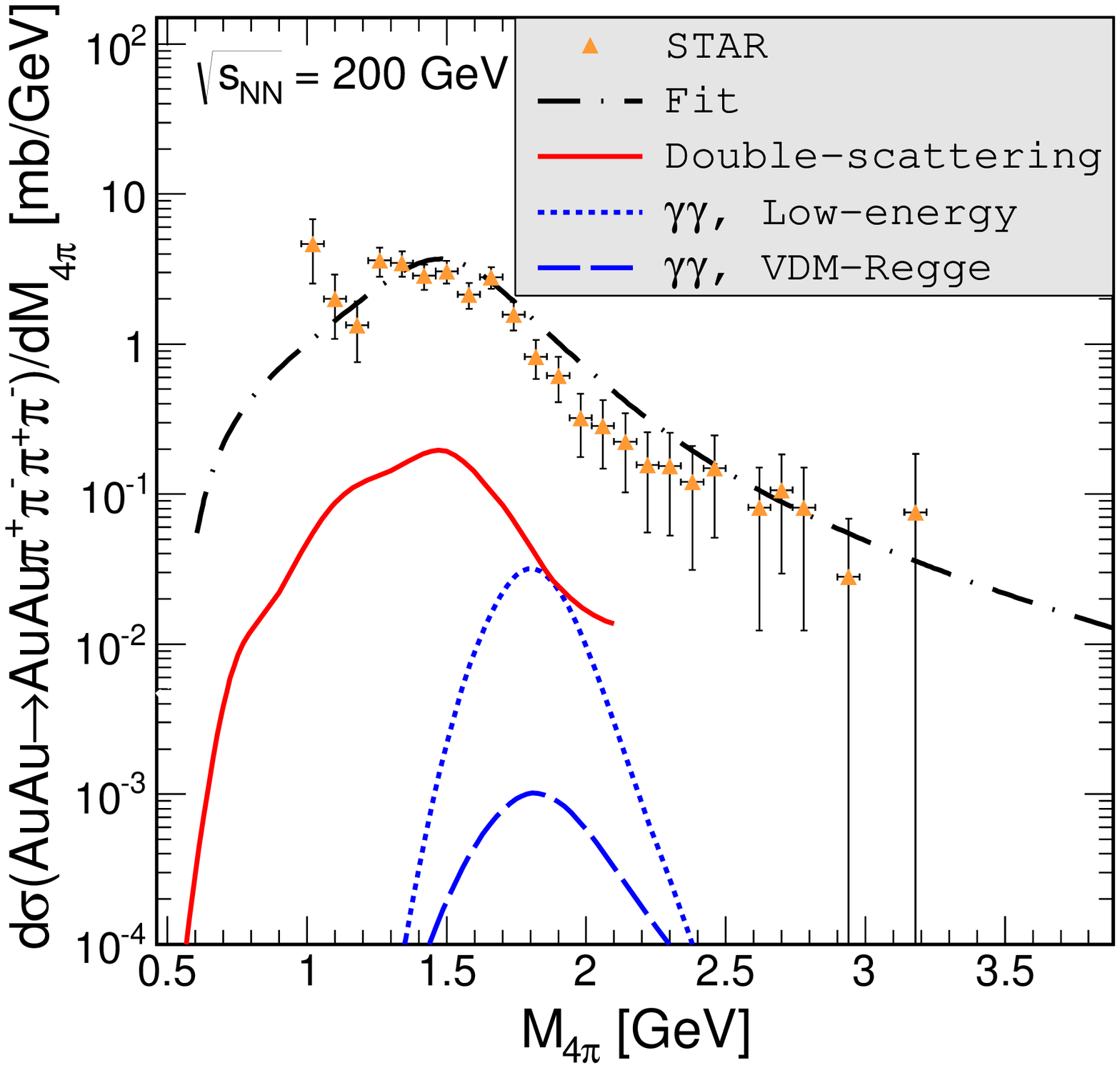}
\includegraphics[scale=0.3]{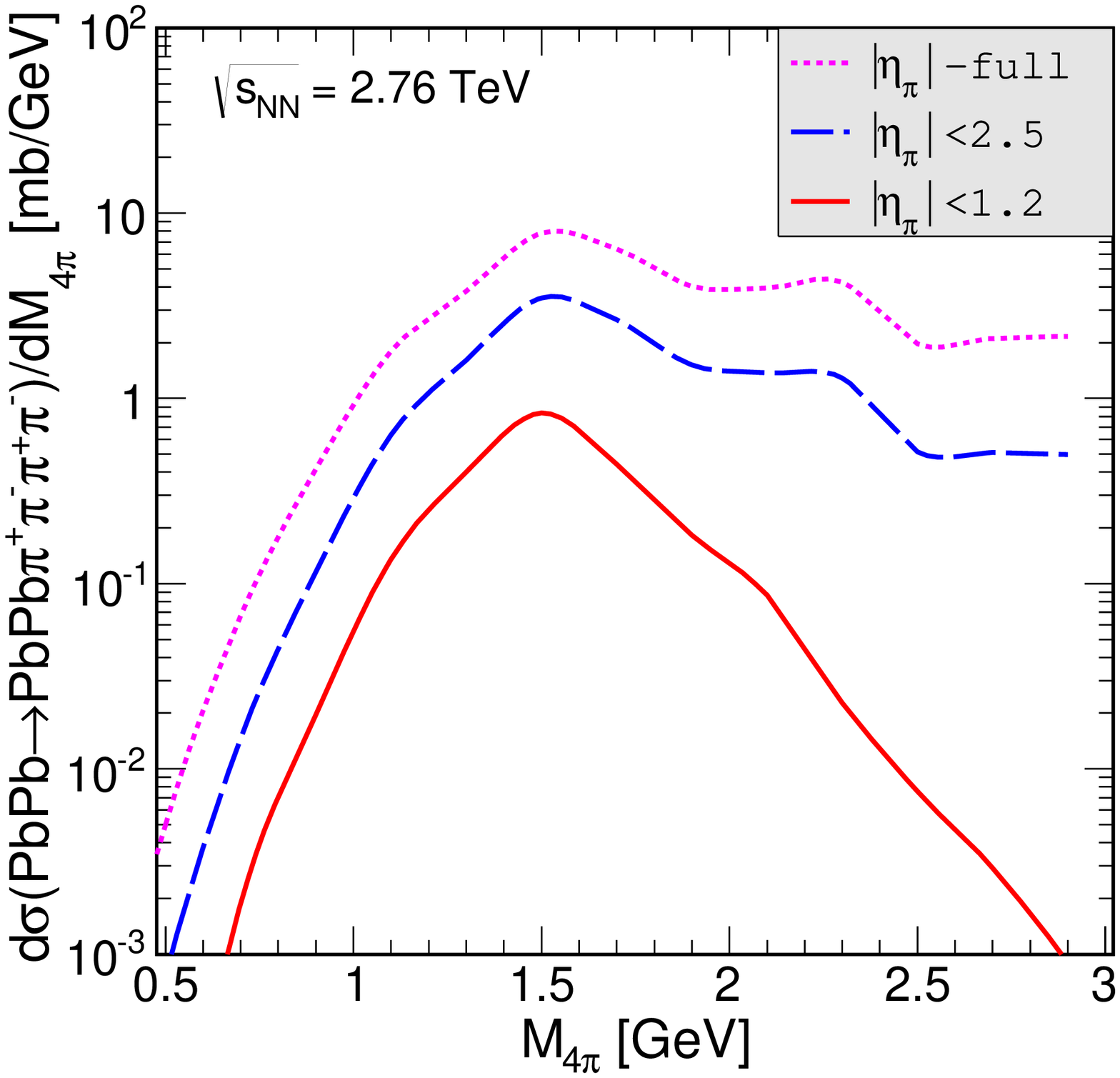}
\caption{Invariant mass of four charged pions produced 
in the double-scattering mechanism and in the $\gamma\gamma$ fusion
of two $\rho^0$ mesons at the RHIC (left panel) and at 
the LHC (right panel) energy.}
\label{fig_dsig_dm4pi}       % Give a unique label
\end{figure}

As in previous figures, Fig. \ref{fig_dsig_dm4pi} shows
contributions from double-scattering, the low-energy bump and the high-energy VDM-Regge 
$\gamma\gamma$ fusion contributions \cite{KSS2009} but now as a function of
four-pion invariant mass. 
The left panel of Fig. \ref{fig_dsig_dm4pi} shows results 
for the limited acceptance of the STAR experiment 
($|\eta_\pi| < 1$) \cite{STAR_4pi}. 
The dash-dotted line represents a fit of the STAR Collaboration. 
The double-scattering contribution accounts only for $20\%$ of the cross section 
measured by the STAR Collaboration. 
The STAR experimental data have been corrected by experimental
acceptance \cite{STAR_4pi}. 
Wu suppose that the production of the $\rho^0(1450)$ or $\rho^0(1700)$ resonance
and their subsequent decay into the four-pion final state is the dominant effect 
for the limited STAR acceptance. However, the production mechanism of
this two broad resonances and their decay into four charged pions are not yet understood. 
A model production of the resonances and their decay has to be study 
in the future.
The right panel of Fig. \ref{fig_dsig_dm4pi} shows four-pion invariant mass 
distribution for double-scattering mechanism for different limited range of 
pion pseudorapidity at the LHC energy. The ALICE group collected
the data for four-charged-pion production with the limitation $|\eta_\pi|<1.2$.
We cannot compare this distribution with the ALICE data,
because those data are not yet absolutely normalized.

\begin{figure}[ht]
\centering
\includegraphics[scale=0.3]{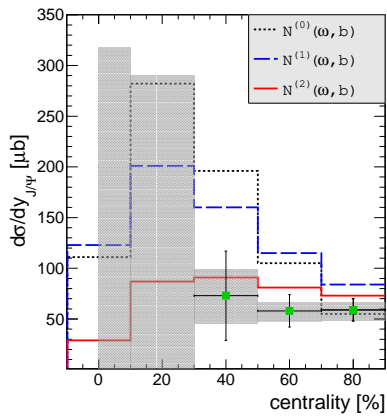}
\caption{$d\sigma/dy$ cross sections for $Pb Pb \to Pb Pb J/\psi$ reaction 
for different centrality bins. 
Theoretical results for three forms of photon fluxes \cite{KGS2016} are compared with 
the ALICE data \cite{ALICE_central}. The shaded area represents the experimental uncertainties.}
\label{fig_dsig_dy_histo}       % Give a unique label
\end{figure}

Going from peripheral to semi-central
collisions we assume that the whole nucleus produces photons
and the photon (or hadronic photon fluctuation) must hit the other nucleus
to produce the vector meson. 
In Fig. \ref{fig_dsig_dy_histo} we show differential nuclear cross section 
for single $J/\psi$ production as a function of centrality for its small values 
($b < R_A + R_B$) i.e. for the semi-central collisions.
We present new experimental ALICE data \cite{ALICE_central} 
with statistical and systematic error bars (shaded area). 
The ALICE Collaboration could not extract actual
values of the cross section for the two lowest centrality bins. 
The results for standard photon flux (see Eq. (2.3) in Ref. \cite{KGS2016})
exceed the ALICE data. Rather good agreement with the data is achieved for
the $N^{(2)}$ photon flux (see Eq. (2.5) in Ref. \cite{KGS2016}) 
obtained with the realistic nucleus form factor.
In this approximation we integrate the photon flux of the first (emitter) nucleus 
only over this part of the second (medium) nucleus which does not
collide with the nucleus-emitter (see right panel of Fig. \ref{fig_per_cent}).

%------------------------
\section{Conclusion}
%------------------------

We have studied one and two vector meson production 
as well as four-pion production in exclusive ultrarelativistic UPC of heavy ions.
In addition, we have explained theoretical transition from ultraperipheral
to peripheral or semi-central cases.
In our calculations we have used equivalent photon approximation
in the impact parameter space. 
We have obtained good 
description of STAR and ALICE UPC data for $\rho^0(770)$ production and
CMS and ALICE data for $J/\psi$ production.
For more central nucleus-nucleus collisions we have calculated, 
for the first time, e.g. the differential cross section as a function
of $J/\psi$ quarkonium rapidity and simultaneously we have obtained very
good agreement with ALICE experimental data. 
We think that correct interpretation of the ALICE data suggests 
that the ”coherent”
(assumed by the formula used for the $\gamma A \to J/\psi A$ process) 
scattering of the hadronic fluctuation happens before  
the quark-gluon plasma is created.
In addition, we have compared contribution of
four-pion production via $\rho^0\rho^0$ production 
(double scattering and $\gamma\gamma$ fusion)
with experimental STAR Collaboration data. The theoretical predictions
have a similar shape as the distribution measured by the STAR Collaboration,
but missing contribution probably come from decays of excited states 
of $\rho^0$ resonances into four charged pions.

\section*{Acknowledgement}
\begin{acknowledgement}
This work was partially supported by the Polish grant No. DEC-2014/15/B/ST2/02528 (OPUS) 
as well as by the Centre for Innovation and Transfer of
Natural Sciences and Engineering Knowledge in Rzesz{\'o}w.
\end{acknowledgement}

%
% BibTeX or Biber users please use (the style is already called in the class, ensure that the "woc.bst" style is in your local directory)
% \bibliography{name or your bibliography database}

\end{document}